\documentclass[runningheads]{llncs}
\usepackage[T1]{fontenc}
\usepackage{amsmath}
\usepackage{amssymb}  

\usepackage{graphicx}
\usepackage{booktabs}
\usepackage{float}
\usepackage{multicol}
\usepackage{multirow}
\usepackage{tikz}
\usetikzlibrary{calc}
\usepackage{subcaption}
\usepackage[misc]{ifsym}
\newcommand{\corr}{(\Letter)}

\usepackage{mwe}
\usepackage{xcolor}

\usepackage{mwe}

\usepackage{xcolor}

\begin{document}

\title{A CNN-based Local-Global Self-Attention via Averaged Window Embeddings for Hierarchical ECG Analysis}

\authorrunning{Buzelin et al.}
\titlerunning{LGA-ECG}

\author{Arthur Buzelin\inst{1} \corr \and
Pedro Robles Dutenhefner\inst{1}  \and
Turi Rezende\inst{1}  \and
Luisa G. Porfirio\inst{1}  \and
Pedro Bento\inst{1}  \and
Yan Aquino\inst{1}  \and
Jose Fernandes\inst{1}  \and
Caio Santana\inst{1}  \and
Gabriela Miana\inst{1}  \and
Gisele L. Pappa\inst{1}  \and
Antonio Ribeiro\inst{1}  \and
Wagner Meira Jr.\inst{1}}



\institute{
Universidade Federal de Minas Gerais, Belo Horizonte, Brazil \\ 
\email{\{buzelin, turirezende, pedro.bento, yanaquino, caiosantana\}@dcc.ufmg.br \\ \{pedroroblesduten, josegeraldof\}@ufmg.br \\
gabimiana@gmail.com, tom@hc.ufmg.br,\\
\{glpappa, meira\}@dcc.ufmg.br}
}

\maketitle              

\begin{abstract}

Cardiovascular diseases remain the leading cause of global mortality, emphasizing the critical need for efficient diagnostic tools such as electrocardiograms (ECGs). Recent advancements in deep learning, particularly transformers, have revolutionized ECG analysis by capturing detailed waveform features as well as global rhythm patterns. However, traditional transformers struggle to effectively capture local morphological features that are critical for accurate ECG interpretation. We propose a novel Local-Global Attention ECG model (LGA-ECG) to address this limitation, integrating convolutional inductive biases with global self-attention mechanisms. Our approach extracts queries by averaging embeddings obtained from overlapping convolutional windows, enabling fine-grained morphological analysis, while simultaneously modeling global context through attention to keys and values derived from the entire sequence. Experiments conducted on the CODE-15 dataset demonstrate that LGA-ECG outperforms state-of-the-art models and ablation studies validate the effectiveness of the local-global attention strategy. By capturing the hierarchical temporal dependencies and morphological patterns in ECG signals, this new design showcases its potential for clinical deployment with robust automated ECG classification. \footnote{Paper under-review}


\end{abstract}

\section{Introduction}

Cardiovascular diseases (CVDs) remain the leading cause of death globally, responsible for 17.9 million deaths in 2019, which corresponds to 32\% of all deaths worldwide, as reported by the World Health Organization (WHO) \cite{who_cardiovascular_diseases}. In light of this, electrocardiograms (ECGs), which are non-invasive and easy-to-perform examinations, are fundamental tools in the detection and monitoring of heart-related conditions. Their importance has only grown with the rise of digital health technologies \cite{liu2021deep}. In this context, artificial intelligence has become a valuable resource for automating ECG analysis, supporting clinical decision-making, reducing backlogs in telemedicine services, and enabling automated tasks such as disease classification \cite{ebrahimi2020review}, age estimation \cite{lima2021deep}, and wave segmentation \cite{fujita2015performance}.\looseness=-1


The rise of deep learning has transformed ECG signal analysis, with deep neural networks (DNNs) excelling in automatic feature extraction from raw ECG data, eliminating the need for manual engineering. Convolutional neural networks (CNNs) are particularly suited for this task, leveraging inductive biases such as spatial locality and translation equivariance \cite{ribeiro2020automatic,rajpurkar2011cardiologist}. These properties enable CNNs to capture hierarchical temporal structures in ECG signals, from localized morphological features within heartbeats to global rhythm patterns. Moreover, translation equivariance ensures robust detection of clinically relevant features, regardless of their temporal positions.

Transformer architectures have shown significant success across various domains. However, the direct application of Vision Transformer (ViT)-based models to ECG data faces limitations due to their global attention mechanisms, which fail to adequately capture localized morphological features essential for accurate ECG interpretation. To address this limitation, hierarchical transformer models integrating local self-attention mechanisms have been proposed, demonstrating superior performance in ECG classification tasks \cite{li2021bat,dutenhefner2024transformer}. These models leverage the strengths of transformers in modeling temporal relationships while preserving locality bias, crucial for ECG signal interpretation. Building on these developments, this paper introduces a novel transformer architecture specifically tailored for ECG data.

Cardiology experts emphasize that effective ECG models must encompass multiple levels of temporal and contextual information, from individual waveform morphology to the overall rhythm structure \cite{dutenhefner2024transformer}. Consequently, combining local feature extraction with global attention mechanisms emerges as a promising strategy for ECG analysis. Motivated by that, we propose a novel hierarchical transformer architecture that leverages overlapping convolutional projections to derive queries from local temporal segments. Specifically, each query vector is computed by averaging convolutional features within overlapping windows, inherently embedding convolutional inductive biases into the self-attention mechanism. These locally informed queries then attend to globally computed key and value vectors, allowing the model to simultaneously capture detailed morphological characteristics (such as waveform shapes and intervals) and broader contextual dependencies (such as inter-beat relationships) within ECG signals.

Our proposed Local-Global Attention ECG model (LGA-ECG) applies local convolutional inductive biases with global self-attention mechanisms, significantly improving ECG classification. Experimental results demonstrate that this hybrid approach outperforms state-of-the-art baseline methods, achieving superior performance in classification tasks.


\section{Related Works}

In this section, we review prior work on deep learning methods for ECG analysis, focusing on CNN-based approaches, hybrid CNN-transformer architectures, and local attention mechanisms; bridging the gap between local feature extraction and global sequence modeling.

\subsection{Neural Networks for ECG}

Automatic analysis of ECG signals has been extensively studied, with recent literature emphasizing deep learning methods, particularly Convolutional Neural Networks (CNNs), due to their inherent capacity for autonomously extracting morphological and temporal ECG features from raw data. Rajpurkar et al.~\cite{rajpurkar2011cardiologist} pioneered a deep CNN approach trained end-to-end on a large-scale dataset, achieving cardiologist-level accuracy in arrhythmia classification. Similarly, Ribeiro et al.~\cite{ribeiro2020automatic} introduced deep neural networks with stacked convolutions, demonstrating strong generalization capabilities across multi-lead ECG data and surpassing cardiologist-level performance. Wang et al.~\cite{wang2011atrial} proposed a CNN-BiLSTM model for atrial fibrillation classification, using CNNs for morphology and BiLSTMs for temporal dependencies in ECG signals.

Recently, transformer-based architectures have increasingly been applied to ECG analysis, motivated by their success in sequence modeling tasks. Both Hu et al.~\cite{hu2022transformer} and El et al.~\cite{el2024ecgtransform} proposed hybrid CNN-Transformer approaches, where convolutional layers initially extract local morphological features, followed by transformer-based self-attention layers that model global temporal interactions. Building on \cite{liu2021swin}, Liu et al.\cite{li2021bat} proposed BaT, a beat-aligned framework leveraging local attention to process ECG signals progressively. BaT segments ECGs into beats, applying self-attention locally before merging representations to capture hierarchical features. However, it depends on complex preprocessing, where inaccuracies in beat segmentation may introduce biases or errors. Additionally, Dutenhefner et al.~\cite{dutenhefner2024transformer} proposed an approach that interleaves CNN and transformer blocks to create a hierarchical, multi-scale feature extraction pipeline.

\subsection{Local Attention}

Local attention mechanisms combine the modeling capabilities of self-attention with the structured inductive biases of convolution, benefiting tasks with spatial and sequential dependencies. Liu et al.~\cite{liu2021swin} introduced the Swin Transformer, which improves efficiency through hierarchical local attention with shifting windows. However, its non-overlapping windows limit direct global context modeling, which is crucial for ECG analysis.

CoAtNet~\cite{dai2021coatnet} addresses this problem by integrating convolutional inductive biases with global transformers via multi-scale relative attention. While effective, its predefined branches reduce flexibility and increase computational complexity. Similarly, Zhou et al.~\cite{zhou2021elsa} introduced ELSA, enhancing local feature extraction with Hadamard attention and a ghost head module.

Building upon these strengths and addressing the limitations identified in previous works, the approach we propose introduces a novel local-global attention mechanism designed for ECG signals. Our method efficiently captures both local morphological variations and long-range dependencies while mitigating the computational burden. By leveraging adaptive attention windows and progressive feature aggregation, our approach also enhances ECG feature representation.

\section{Methods}

ECG analysis requires capturing information across multiple temporal scales: wave morphology (P, QRS, T), intra-heartbeat intervals (PR, QT), and inter-beat distances essential for rhythm analysis. We propose a novel self-attention mechanism tailored for ECG signals, which effectively balances fine-grained morphological details with global heartbeat patterns.

The proposed model first uses convolutional layers to project the ECG into an embedding space. Its core comprises layers of a novel windowed self-attention and feed-forward blocks with residual connections. Unlike traditional global self-attention, our method extracts queries (Q) from small overlapping windows to preserve local detail, while keys (K) and values (V) are computed globally, capturing long-range dependencies.
Additionally, each self-attention block progressively reduces the sequence length, similar to convolutional pooling, allowing hierarchical abstraction from local waveform characteristics toward global rhythm and beat-to-beat features.


\subsection{Local-Global Self-Attention}


The core innovation of our proposed transformer-based architecture lies in its novel local-global self-attention mechanism. Traditional self-attention mechanisms compute interactions uniformly across all tokens, which may overlook crucial local patterns in biomedical signals. In contrast, our method balances fine-grained local feature extraction and broader temporal context modeling.

Let us formally define the input tensor to this attention mechanism as \(\mathbf{X}\), with dimensions:
\begin{equation}
\mathbf{X} \in \mathbb{R}^{B \times N \times D},
\end{equation}
where \(B\) is the batch size, \(N\) the sequence length, and \(D\) the embedding dimension.

\textbf{Step 1: Normalization}.  
First, we apply a standard layer normalization along the embedding dimension to stabilize and normalize the input:
\begin{equation}
\tilde{\mathbf{X}} = \text{LayerNorm}(\mathbf{X}), \quad \tilde{\mathbf{X}} \in \mathbb{R}^{B \times N \times D}.
\end{equation}

\textbf{Step 2: Local Windowed Query Generation}.
To effectively capture precise wave-level morphological details from ECG signals, we introduce a local window-based query generation strategy. Starting from the normalized input tensor \(\tilde{\mathbf{X}} \in \mathbb{R}^{B \times N \times D}\), we extract a series of overlapping windows along the temporal dimension to form localized queries (\(\mathbf{Q}\)). 

Formally, given a window length \( l \) and stride \( s \), we extract \( M \) overlapping windows from the sequence, where:
\begin{equation}
M = \left\lfloor \frac{N - l}{s} \right\rfloor + 1.
\end{equation}

For each window indexed by \( i \in \{0, 1, \dots, M-1\} \), we select a contiguous subset of the input sequence:
\begin{equation}
\tilde{\mathbf{X}}^{(i)} = \tilde{\mathbf{X}}\left[:,\, (i \cdot s):(i \cdot s + w),\,:\right],\quad \tilde{\mathbf{X}}^{(i)} \in \mathbb{R}^{B \times w \times D}.
\end{equation}

Next, each extracted window \(\tilde{\mathbf{X}}^{(i)}\) undergoes a convolutional projection along the temporal dimension. Specifically, we apply a 1D convolution with kernel size \(k_q\), stride \(1\), padding \(p_q\), and \(D\) output channels, obtaining:
\begin{equation}
\mathbf{Q}^{(i)}_{\text{conv}} = \text{Conv1D}_Q\left(\tilde{\mathbf{X}}^{(i)}\right),
\quad \mathbf{Q}^{(i)}_{\text{conv}} \in \mathbb{R}^{B \times D \times w}.
\end{equation}

The output \(\mathbf{Q}^{(i)}_{\text{conv}}\) represents an enhanced embedding of the original local window, where each temporal position within the window has been projected into a new feature space through convolution.

To summarize this detailed local information into a single representative query vector per window, we then average these embeddings along the temporal dimension of length \(w\). For each window \(i\), the averaged query vector is calculated as:
\begin{equation}
\mathbf{Q}^{(i)} = \frac{1}{w}\sum_{t=1}^{w}\mathbf{Q}^{(i)}_{\text{conv}}[:, :, t],
\quad \mathbf{Q}^{(i)} \in \mathbb{R}^{B \times D}.
\end{equation}

Finally, stacking all the averaged queries across the \(M\) extracted windows results in the complete query tensor for attention:
\begin{equation}
\mathbf{Q} = \left[\,\mathbf{Q}^{(0)},\, \mathbf{Q}^{(1)},\, \dots,\, \mathbf{Q}^{(M-1)}\,\right],
\quad \mathbf{Q} \in \mathbb{R}^{B \times M \times D}.
\end{equation}

\begin{figure}[ht]
    \centering
    \includegraphics[width=0.52\linewidth]{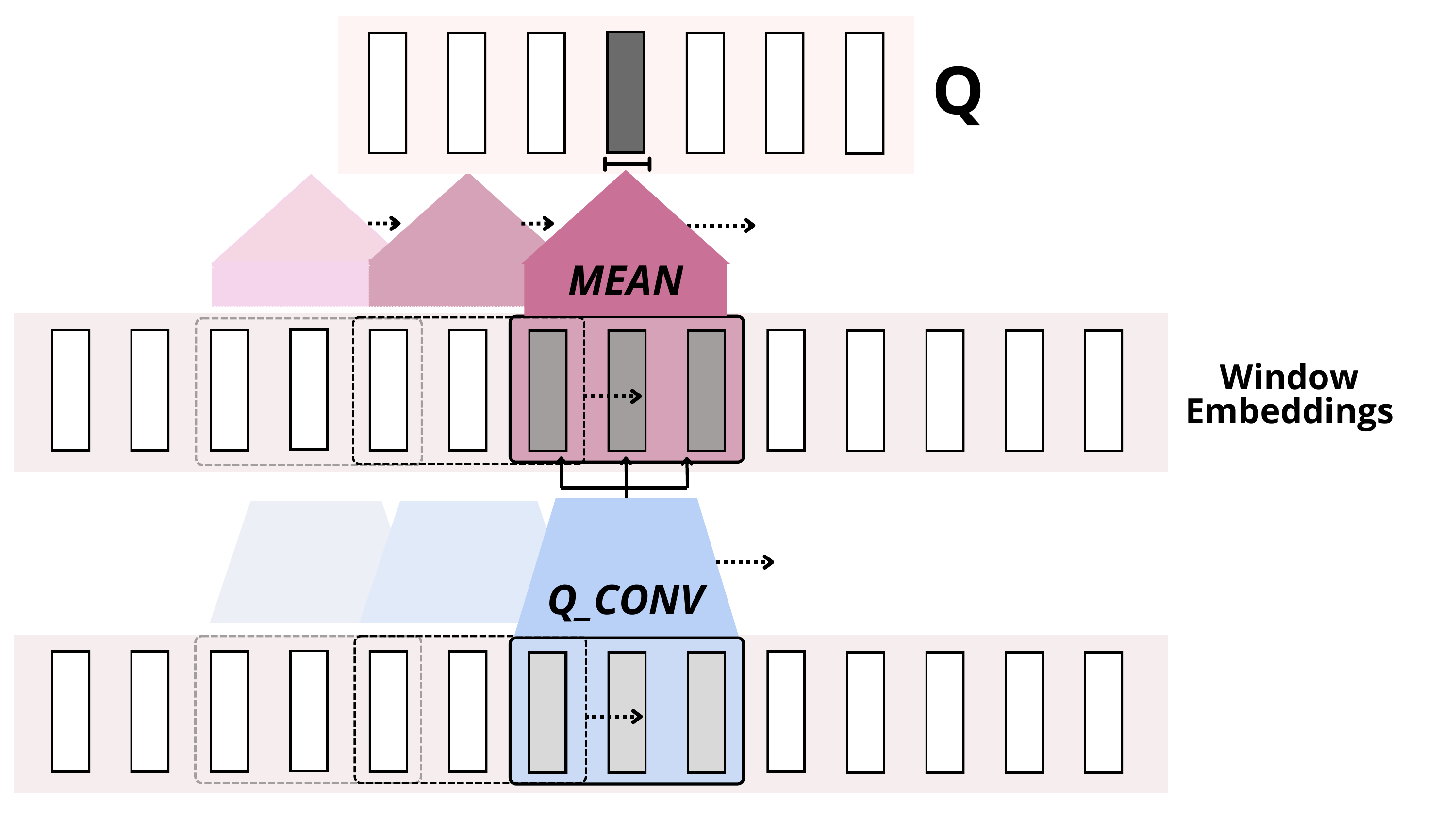}
    \caption{Mean query extraction process for each ECG window.}
    \label{fig:q_mean}
\end{figure}

To enhance stability and facilitate residual connections in deeper layers, we retain a copy of the query tensor as a residual term. This preserves local morphological details captured by convolution, ensuring stable gradients and improved convergence.

This process can be implemented in a simple and effective manner using a combination of a 1D convolutional layer that preserves the input shape, followed by an average pooling layer. The kernel size of the pooling operation determines the temporal compression factor. This approach is illustrated in Figure \ref{fig:q_mean}.

\textbf{Step 3: Global Key and Value Generation}.  
In contrast to the localized queries, keys (\(\mathbf{K}\)) and values (\(\mathbf{V}\)) are computed from the entire normalized sequence, enabling each local query to attend globally. We define these global projections using convolutional layers to retain a locality inductive bias while still allowing global context modeling:
\begin{equation}
\mathbf{K}_{\text{conv}} = \text{Conv1D}_K(\tilde{\mathbf{X}}), \quad \mathbf{V}_{\text{conv}} = \text{Conv1D}_V(\tilde{\mathbf{X}}),
\end{equation}
both producing tensors of shape:
\begin{equation}
\mathbf{K}_{\text{conv}}, \mathbf{V}_{\text{conv}} \in \mathbb{R}^{B \times D \times N}.
\end{equation}

We permute them back to match the original embedding format:
\begin{equation}
\mathbf{K} = \mathbf{K}_{\text{conv}}^\top \in \mathbb{R}^{B \times N \times D}, \quad
\mathbf{V} = \mathbf{V}_{\text{conv}}^\top \in \mathbb{R}^{B \times N \times D}.
\end{equation}

\textbf{Step 4: Multi-Head Local-Global (LG) Attention Computation}.  
We now apply a multi-head attention mechanism. For \(H\) attention heads, we split the embedding dimension \(D\) into \(H\) sub-dimensions of size \(D_h = D/H\):
\begin{equation}
\mathbf{Q}_h \in \mathbb{R}^{B \times M \times D_h},\quad
\mathbf{K}_h,\mathbf{V}_h \in \mathbb{R}^{B \times N \times D_h},\quad h=1,\dots,H.
\end{equation}

For each head \(h\), the scaled dot-product attention scores are computed as:
\begin{equation}
\mathbf{A}_h = \text{softmax}\left(\frac{\mathbf{Q}_h \mathbf{K}_h^\top}{\sqrt{D_h}}\right) \in \mathbb{R}^{B \times M \times N}.
\end{equation}

Subsequently, we calculate the features as a weighted sum of values:
\begin{equation}
\mathbf{O}_h = \mathbf{A}_h\mathbf{V}_h \in \mathbb{R}^{B \times M \times D_h}.
\end{equation}

Concatenating across all heads, we get the combined multi-head attention output:
\begin{equation}
\mathbf{O} = \text{concat}(\mathbf{O}_1,\dots,\mathbf{O}_H) \in \mathbb{R}^{B \times M \times D}.
\end{equation}

\textbf{Step 5: Residual Connection and Sequence Reduction}.  
Finally, we reintroduce the residual query information by adding back the previously stored queries \(\mathbf{Q}_{\text{res}}\), maintaining strong local fidelity:
\begin{equation}
\mathbf{Y} = \mathbf{O} + \mathbf{Q}_{\text{res}},\quad \mathbf{Y}\in \mathbb{R}^{B \times M \times D}.
\end{equation}
The sequence length is effectively reduced from \(N\) to \(M\) by selecting a stride \(s = 2\), ensuring \(M = N/2\). This hierarchical summarization progressively condenses ECG features, capturing local and global information.

\begin{figure}[h]
    \centering
    \includegraphics[width=0.75\linewidth]{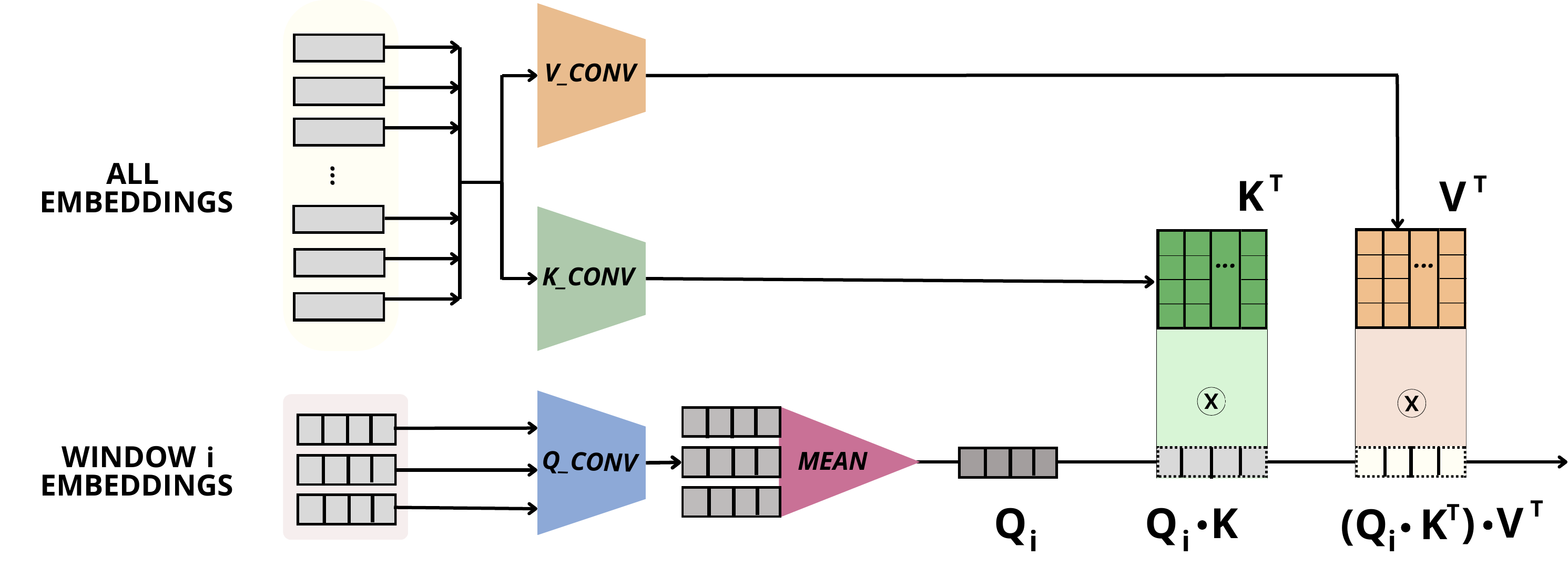}
    \caption{Local-global self-attention operation for one ECG embedding window.}
    \label{fig:attention}
\end{figure}

Our LG self-attention combines standard self-attention, convolution, and hierarchical transformers while overcoming their limitations. Unlike traditional self-attention, which lacks locality and scales quadratically, or convolutions, which struggle with long-range dependencies, our method extracts locally-informed queries via overlapping convolutional projections while maintaining global attention through sequence-wide keys and values. Additionally, convolutional projections inherently encode positional information, removing the need for explicit positional encodings. The local-global attentions is illustrated in Figure \ref{fig:attention}

\subsection{Transformer Block with Local-Global Self-Attention}

The Transformer Block integrates the local-global self-attention mechanism within a standard transformer architecture. It consists of two sub-layers: LGA and a feed-forward network, both with layer normalization and residual connections for stable training.

Given an input tensor \(\mathbf{X} \in \mathbb{R}^{B \times N \times D}\), where \(B\) is the batch size, \(N\) is the sequence length, and \(D\) is the embedding dimension, the Transformer Block initially applies layer normalization along the embedding dimension:

\begin{equation}
\tilde{\mathbf{X}} = \text{LayerNorm}(\mathbf{X}), \quad \tilde{\mathbf{X}} \in \mathbb{R}^{B \times N \times D}.
\end{equation}

Subsequently, the normalized sequence is processed by the local-global self-attention layer. Due to the windowed attention design, the spatial dimension \(N\) is effectively reduced approximately by half, from \(N\) to \(M = N/2\), resulting in an output tensor \(\mathbf{Y}_{\text{attn}}\):

\begin{equation}
\mathbf{Y}_{\text{attn}} = \text{LocalGlobalAttention}(\tilde{\mathbf{X}}), \quad \mathbf{Y}_{\text{attn}} \in \mathbb{R}^{B \times M \times D}.
\end{equation}

To maintain a consistent residual connection despite the reduction in sequence length, we apply a pooling operation followed by a \(1\times1\) convolution to the normalized input \(\tilde{\mathbf{X}}\), ensuring dimensional compatibility:

\begin{equation}
\mathbf{X}_{\text{res}} = \text{Conv1D}\left(\text{MaxPool1D}\left(\tilde{\mathbf{X}}\right)\right), \quad \mathbf{X}_{\text{res}} \in \mathbb{R}^{B \times M \times D}.
\end{equation}

Here, the max pooling operation reduces the temporal dimension by half, from \(N\) to \(M\), while the \(1\times 1\) convolution adjusts embedding dimensions and reinforces the residual pathway. The resulting residual tensor \(\mathbf{X}_{\text{res}}\) is added to the self-attention output, stabilizing training and enhancing gradient flow:

\begin{equation}
\mathbf{Z} = \mathbf{Y}_{\text{attn}} + \mathbf{X}_{\text{res}}, \quad \mathbf{Z} \in \mathbb{R}^{B \times M \times D}.
\end{equation}

Next, we apply a second-layer normalization followed by a feed-forward neural network, often called the Multi-Layer Perceptron (MLP). This MLP consists of two linear layers with an intermediate non-linearity (\(\text{ReLU}\)). The dimensionality of the intermediate MLP layer, denoted as \(D_{\text{MLP}}\), dynamically increases at each transformer block stage \(i\), defined explicitly as \(D_{\text{MLP}} = D_{\text{base}} \times 2 \times i\). Specifically, the MLP initially projects each embedding vector from the input dimension \(D\) to this expanded dimension \(D_{\text{MLP}}\):

\begin{equation}
\mathbf{Z}_{\text{MLP}}^{(i)} = \text{ReLU}\left(\mathbf{Z}^{(i)}\mathbf{W}_1^{(i)} + \mathbf{b}_1^{(i)}\right), \quad \mathbf{Z}_{\text{MLP}}^{(i)} \in \mathbb{R}^{B \times M \times (D_{\text{base}} \times 2 \times i)},
\end{equation}

and subsequently project it back to the original embedding dimension \(D\):

\begin{equation}
\mathbf{Z}_{\text{out}}^{(i)} = \mathbf{Z}_{\text{MLP}}^{(i)}\mathbf{W}_2^{(i)} + \mathbf{b}_2^{(i)}, \quad \mathbf{Z}_{\text{out}}^{(i)} \in \mathbb{R}^{B \times M \times D}.
\end{equation}

This incremental expansion of the MLP dimensionality at successive transformer stages allows the model to progressively capture more complex and abstract features. A second residual connection then integrates the MLP output back into the main pathway, resulting in the final output tensor of each transformer block:

\begin{equation}
\mathbf{X}_{\text{final}}^{(i)} = \mathbf{Z}^{(i)} + \mathbf{Z}_{\text{out}}^{(i)}, \quad \mathbf{X}_{\text{final}}^{(i)} \in \mathbb{R}^{B \times M \times D}.
\end{equation}

This staged expansion of the MLP dimension allows deeper layers to encode increasingly complex and abstract features, naturally aligning with the progressive shift from fine-grained morphological details to broader, long-range inter-beat relationships.

Each Transformer Block hierarchically condenses and enriches representations, aligning with clinical ECG analysis. Early layers capture fine-grained wave morphology, intermediate layers focus on intra-heartbeat intervals, and deeper layers model long-range dependencies across heartbeats, effectively identifying rhythm abnormalities. This structured progression inherently encodes clinically relevant inductive biases.

\subsection{Overall Model Architecture}

\begin{figure}[h]
    \centering
    \includegraphics[width=0.8\linewidth]{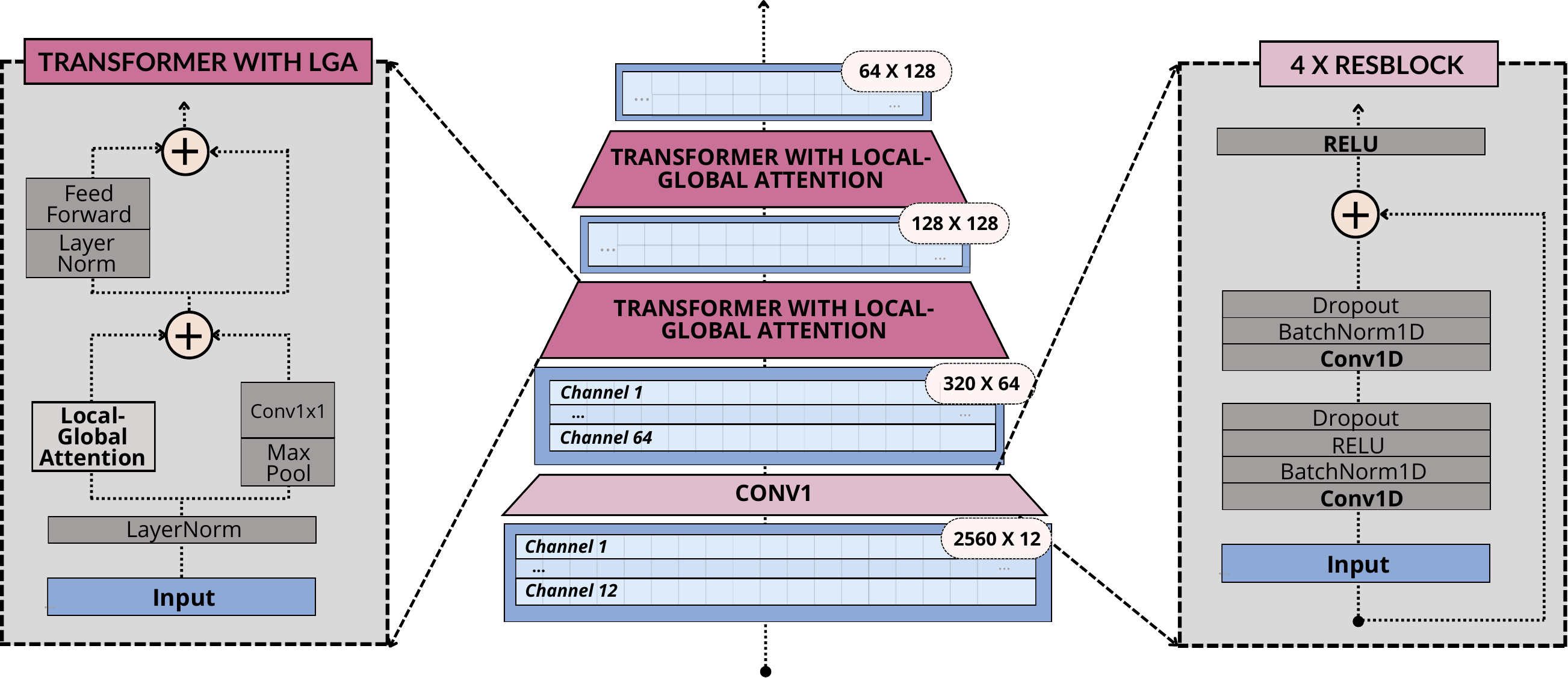}
    \caption{Overall architecture of the proposed LGA network, integrating the convolutional front-end, composed of four repeated ResBlocks (right), with transformer blocks utilizing local-global self-attention (left).}
    \label{fig:architecture}
\end{figure}

The overall architecture of the proposed model is illustrated in Figure~\ref{fig:architecture}. It comprises two main components: a convolutional fron-tend and a sequence of transformer blocks equipped with the LGA mechanism. Initially, the multi-scale convolutional front-end transforms the raw ECG signals into a sequence of feature embeddings, capturing localized waveform details while reducing temporal dimensions. Subsequently, these embeddings are processed by a cascade of transformer blocks featuring the proposed local-global self-attention. These blocks hierarchically aggregate ECG features at progressively coarser temporal scales, effectively encoding wave-level morphologies, intra-beat intervals, and inter-beat rhythm relationships into comprehensive representations suitable for ECG analysis.

\section{Experiment setup}
\subsection{Datasets}
Our model was trained and evaluated using CODE-15, a publicly available 15\% subset of the CODE (Clinical Outcomes in Digital Electrocardiography ) dataset \cite{ribeiro2019tele}. CODE contains over 2 million ECGs from Minas Gerais, Brazil, annotated by cardiologists for six cardiac abnormalities: first-degree atrioventricular block (1st AVB), right bundle branch block (RBBB), left bundle branch block (LBBB), sinus bradycardia (SB), atrial fibrillation (AF), and sinus tachycardia (ST). These conditions indicate an increased risk for cardiovascular events, including stroke, heart failure, and sudden death, and require targeted clinical interventions. CODE-15 comprises 345,779 exams from 233,770 patients and has been widely adopted in ECG research, serving as a benchmark dataset for developing and evaluating deep learning models \cite{ribeiro2020automatic} \cite{tulerleveraging}.

We evaluated our model using the publicly available CODE-TEST dataset, also collected by the Telehealth Network of Minas Gerais (TNMG). CODE-TEST comprises 827 ECGs labeled by consensus among two or three cardiologists, covering the same six cardiac abnormalities. The high-quality, expert-consensus labels provide a robust benchmark for performance assessment.

For developing and validating the LGA-ECG model, \noindent the dataset is divided into four subsets by patient IDs: 90\% of CODE-15 is used as the training set to train the model, while 5\% of CODE-15 serves as the validation set for early stopping. An additional 5\% of CODE-15 is designated as the development set, which is utilized for hyperparameter tuning and ablation studies. Finally, the entire CODE-TEST dataset is used as the test set to evaluate the final model performance against baseline methods.

\subsection{Implementation details and Benchmarks}
For comparison, we assessed LGA-ECG against a suite of baseline models spanning diverse architectural families, including traditional CNN and transformer-based architectures. 
This selection ensured a rigorous and comprehensive evaluation across distinct modeling paradigms. The baselines were implemented using their original authors' codebases, with training settings configured according to their recommendations. All models were trained on the same Training Set and evaluated on the Test Set to ensure consistent comparisons. We employed standard classification metrics to evaluate the models: accuracy, F1-score, precision, and recall. These metrics were computed for each cardiac condition individually to provide a detailed understanding of model performance across different diseases, as well as averaged (macro).

The training process utilized the AdamW optimizer \cite{loshchilov2017decoupled} and employed a cosine annealing learning rate schedule \cite{loshchilov2016sgdr}. The initial learning rate was set to 0.0001 and was decreased cosine-wise to 0.00001 throughout the training. Additionally, early stopping was implemented, which terminates training if the validation error does not decrease for seven consecutive epochs. The training was conducted in parallel using 4 NVIDIA V100 GPUs.

\section{Results}

In this section, we analyze our model by comparing its performance to state-of-the-art models in ECG abnormality classification.

\begin{table}
\centering
\caption{Average performance of LGA-ECG when compared to other SOTA methods.}
\label{tab:performance_metrics}
\resizebox{\textwidth}{!}{%
\begin{tabular}{lccccccc}
\toprule
\textbf{Metrics} & \textbf{| ResNet-1} & \textbf{| ResNet-2} & \textbf{| ECG-Transform} & \textbf{| BAT} & \textbf{| ECG-DETR} & \textbf{| HiT}  & \textbf{| LGA-ECG}\\
\midrule
\textbf{Accuracy}  & 0.991 & 0.989 & 0.981 & 0.991 & 0.9842 & 0.991 & \textbf{0.994}\\
\textbf{Precision}  & 0.875 & 0.908 & 0.711 & \textbf{0.918} & 0.7768 & 0.909 & 0.907\\
\textbf{Recall}    & 0.778 & 0.743 & 0.687 & 0.799 & 0.6614 & 0.798 & \textbf{0.872}\\
\textbf{F1-Score}  & 0.814 & 0.811 & 0.677 & 0.848 & 0.699 & 0.841 & \textbf{0.885}\\
\bottomrule
\end{tabular}
}
\end{table}

We first compare our model, LGA-ECG, with state-of-the-art methods for classifying six ECG abnormalities: AVB, RBBB, LBBB, SB, AF, and ST. The evaluated models include ResNet-1 \cite{ribeiro2020automatic}, ResNet-2 ~\cite{resnet-2}, BAT \cite{li2021bat}, ECG-DETR \cite{hu2022transformer}, and HiT \cite{dutenhefner2024transformer}. Due to class imbalance, accuracy may be inflated; thus, recall, precision, and particularly the F1-score will be our primary comparison metrics.

Analyzing recall first, LGA-ECG achieves 0.862, surpassing BAT (0.799) and becoming the first to exceed the 0.8 threshold. This marks a significant improvement in identifying positive cases, which is particularly critical in medical applications, where missing a diagnosis due to low recall (high false negatives) can lead to delayed treatments and severe consequences for patients. However, despite this substantial increase in recall, our model maintains a competitive precision of 0.907 -- only slightly lower than BAT’s 0.918. This demonstrates that the boost in recall did not come at the cost of a drastic drop in precision, ensuring a balanced performance that enhances overall reliability. This balance underscores its robustness for practical deployment in high-stakes scenarios.


\begin{table}
\centering
\caption{Per class f1-score of LGA-ECG and baseline methods in the test set.}
\label{tab:f1_results}
\resizebox{\textwidth}{!}{%
\begin{tabular}{lccccccc}
\toprule
\textbf{Abnormality} & \textbf{| ResNet-1} & \textbf{| ResNet-2} & \textbf{| ECG-Transform} & \textbf{| BAT} & \textbf{| ECG-DETR} & \textbf{| HiT} & \textbf{| LGA-ECG} \\
\midrule
1st AVB   & 0.661 & 0.719 & 0.489 & 0.689 & 0.631 & 0.682 & \textbf{0.8}\\
RBBB       & \textbf{0.924}& 0.890 & 0.909 & 0.922 & 0.747 & 0.886 & 0.923\\
LBBB       & 0.927 & 0.843 & 0.886 & 0.945 & 0.826 & 0.909 & \textbf{0.983}\\
SB         & 0.767 & 0.821 & 0.535 & \textbf{0.836} & 0.588 & 0.824 & 0.778\\
AF         & 0.703 & 0.758 & 0.478 & 0.818 & 0.563 & 0.833 & \textbf{0.880}\\
ST         & 0.897 & 0.833 & 0.763 & 0.870 & 0.838 & 0.914 & \textbf{0.946}\\
\midrule
\textbf{Avg. F1} & 0.814 & 0.811 & 0.677 & 0.848 & 0.699 & 0.841 & \textbf{0.885}\\
\bottomrule
\end{tabular}
}
\end{table}

Beyond recall and precision, F1-score provides a comprehensive measure of performance by balancing both metrics. LGA-ECG achieves a new record F1-score of 0.885, surpassing BAT, which reached 0.848, ensuring robust classification across all ECG abnormalities. Despite the class imbalance, our model also achieves a higher accuracy, scoring 0.994 compared to 0.991 from the closest competitor (BAT). By outperforming all baseline methods in all metrics, our approach demonstrates superior overall performance in distinguishing abnormal ECG patterns. All comparisons can be seen in Table \ref{tab:performance_metrics}.

Now, focusing solely on the F1-score, we can directly compare performance across different abnormalities. As shown in Table \ref{tab:f1_results}, LGA-ECG outperforms the baselines in four categories: ST, LBBB, AF, and 1st AVB. For RBBB, although the F1-score is slightly lower, it remains virtually equivalent to that of ResNet-1. The only class in which our model underperforms is SB. We hypothesize that this occurs due to the model's difficulty in accurately detecting longer intervals between consecutive R peaks (the prominent upward deflections in the ECG that indicate ventricular contractions). Further evaluation is necessary to confirm this limitation and guide appropriate improvements.

\begin{figure*}
    \centering
    \includegraphics[width=1.\linewidth]{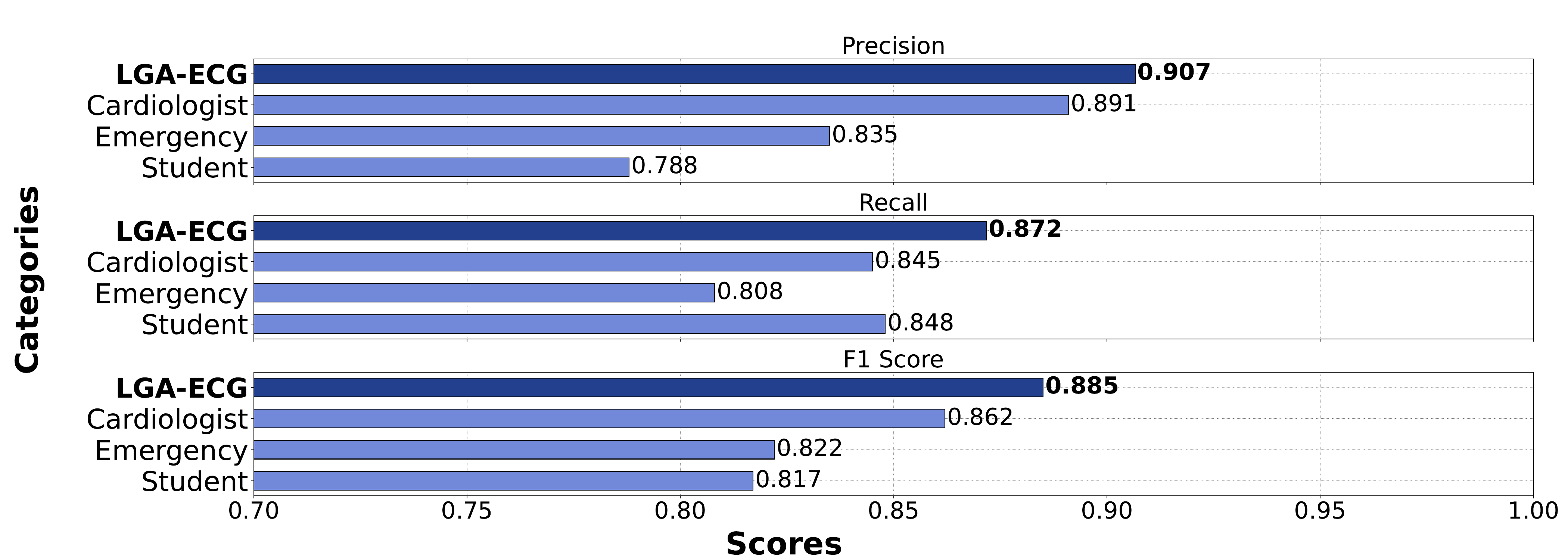}
    \caption{Comparison of the average Precision, Recall, and
F1 Score between the proposed LGA-ECG model and human performance.}
    \label{fig:medico}
\end{figure*}

The CODE-TEST dataset includes labels provided by cardiologists at different levels of training: (i) 4th-year cardiology residents (cardio.), (ii) 3rd-year emergency residents (emerg.), and (iii) 5th-year medical students (stud.). The reference labels used for calculating the evaluation metrics were determined by consensus among three experienced specialist cardiologists, who were excluded from the analysis. Using this expert consensus as the ground truth, we compared the labels assigned by professionals at different training levels with the predictions generated by the proposed LGA-ECG model. The results, presented in Figure \ref{fig:medico}, show that the model outperformed all groups of cardiologists across all key metrics, demonstrating superior performance compared to individuals with varying levels of cardiology expertise.


\section{Ablations}

To assess the effectiveness of our proposed local-global attention mechanism, we perform a series of ablation studies to isolate its contributions and better understand its impact on ECG feature extraction.

\subsection{Alternative Attention Mechanisms}
First, we compare the proposed LGA against alternative attention strategies. Our goal is to evaluate how different query, key, and value configurations influence the model’s ability to capture fine-grained ECG morphology and global contextual dependencies.

\textbf{ViT-like:} We begin by examining a standard ViT-like approach, which applies global self-attention across the entire sequence using linear projections for queries, keys, and values. While this method captures the global context effectively, it lacks local inductive biases.

\textbf{Swin-like:} Next, we compare our method with a local attention mechanism inspired by Swin Transformer~\cite{liu2021swin}, where self-attention is restricted to non-overlapping windows. This approach captures local features while progressively integrating global context through stacked local attention and inter-block pooling.

\textbf{Global Q, K, V: } We also analyze a global attention variant, which follows the standard attention mechanism but replaces linear projections with convolutional and average pooling layers. In this configuration, queries are computed in the same manner as keys and values, ensuring that all positions attend to each other globally. Although this setup preserves global context awareness, it may fail to efficiently encode localized waveform structures.

\textbf{Local Q, K, V: } Finally, we examine a fully localized variant, where the query \( Q \) is the mean of the embeddings within a window, while the keys \( K \) and values \( V \) correspond only to the embeddings of that window, without global context. We extract overlapping windows, ensuring that each window is condensed into a single embedding after the attention operation. This progressively reduces the data by half at each stage, establishing a hierarchical processing framework.

\begin{table}[h]
\centering
\caption{Per class F1-score comparison between different attention mechanisms.}
\label{tab:ablations_results}
\setlength{\tabcolsep}{1pt} 
\renewcommand{\arraystretch}{0.7} 
\scriptsize 
\begin{tabular}{lccccc}
\toprule
\textbf{Abnormality} & \textbf{| ViT-like} & \textbf{| Swin-like} & \textbf{| Global Q, K, V} & \textbf{| Local Q, K, V} & \textbf{| LGA-ECG} \\
\midrule
1st AVB   & 0.653 & 0.682 & 0.809 & 0.782 & \textbf{0.800} \\
RBBB      & 0.862 & 0.886 & 0.925 &\textbf{0.955}& 0.923 \\
LBBB      & 0.875 & 0.909 & 0.909 & 0.982 & \textbf{0.983} \\
SB        & 0.768 & \textbf{0.824} & 0.733 & 0.750 & 0.778 \\
AF        & 0.792 & 0.833 & 0.833 & 0.782 & \textbf{0.880} \\
ST        & 0.887 & 0.914 & 0.870 & 0.885 & \textbf{0.946} \\
\midrule
\textbf{Avg. F1} & 0.806 & 0.841 & 0.847 & 0.856 & \textbf{0.885} \\
\bottomrule
\end{tabular}
\end{table}

\noindent The results in Table~\ref{tab:ablations_results} show that LGA-ECG achieves the highest F1-score (0.885), outperforming all alternative attention mechanisms. By integrating local convolutional inductive biases with global context, LGA-ECG surpasses both fully global (ViT-like, global QKV) and fully local (Swin-like, local QKV) approaches, demonstrating superior feature extraction for ECG classification. 

\subsection{Positional Encoding}

We further evaluate whether convolutional biases introduced by the adapted projections sufficiently capture positional information, which is crucial in ECG analysis due to the diagnostic relevance of intervals between waves and heartbeats. Specifically, we investigate three positional encoding strategies:

\textbf{Absolute sinusoidal positional encoding:} Predefined sinusoidal functions of varying frequencies are computed based on absolute positions and directly summed to the embeddings after the convolutional projection, explicitly embedding absolute positional information into each token.

\textbf{Absolute learnable positional encoding:} A trainable embedding vector for each absolute position is learned during training and summed to the embeddings immediately after convolutional projection, enabling the model to adaptively capture position-specific patterns.

\textbf{Relative positional encoding:} A learnable relative position matrix, matching the attention matrix dimensions, is added directly to the attention scores before the softmax operation. This matrix encodes pairwise relative distances between token positions, allowing the model to flexibly emphasize or suppress interactions based on relative position.

\begin{table}[h]
\centering
\caption{Per class F1-score comparison between positional encoding strategies.}
\label{tab:pe_ablation_results}
\setlength{\tabcolsep}{1pt} 
\renewcommand{\arraystretch}{0.7} 
\scriptsize 
\begin{tabular}{lcccc}
\toprule
\textbf{Abnormality} & \textbf{| Sinusoidal APE} & \textbf{| Learnable APE} & \textbf{| RPE} & \textbf{| Without PE} \\
\midrule
1st AVB & 0.681 & 0.526 & 0.667 & \textbf{0.800} \\
RBBB    & 0.857 & 0.844 & 0.928 & \textbf{0.923} \\
LBBB    & 0.966 & 0.947 & 0.909 & \textbf{0.983} \\
SB      & 0.743 & 0.643 & \textbf{0.800} & 0.778 \\
AF      & 0.769 & 0.667 & 0.621 & \textbf{0.880} \\
ST      & 0.873 & 0.899 & 0.853 & \textbf{0.946} \\
\midrule
\textbf{Avg. F1} & 0.815 & 0.754 & 0.796 & \textbf{0.885} \\
\bottomrule
\end{tabular}
\end{table}

\noindent The results in Table~\ref{tab:pe_ablation_results} indicate that LGA-ECG achieves the highest performance without explicit positional encoding, suggesting that the convolutional projections effectively encode spatial dependencies inherent in ECG signals. While relative positional encoding improves certain classes, neither absolute nor relative positional encodings consistently enhance performance, reinforcing the effectiveness of the learned convolutional inductive biases in capturing diagnostic temporal structures. Notably, relative positional encoding (RPE) improved SB detection, likely aiding R-R interval analysis for bradycardia and rhythm abnormalities. A similar trend in the Swin-like attention, which also uses RPE, highlights its role in enhancing rhythm irregularity detection.

\subsection{Window Size Analysis}

We investigate the impact of varying the window size on the proposed LGA-ECG architecture. This hyperparameter controls both the kernel size of convolutional projections and the temporal length of local segments used to compute the local queries. By testing different window sizes, we aim to evaluate the sensitivity of the model's performance to the temporal scale at which local morphological features are captured.

\begin{figure*}
    \centering
    \includegraphics[width=0.9\linewidth]{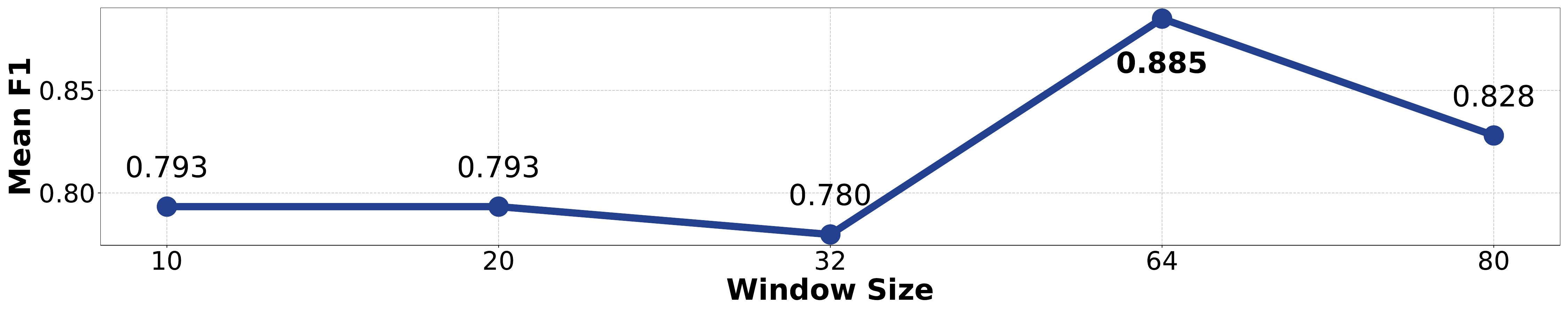}
    \caption{F1-score comparison across different window sizes.}
    \label{fig:enter-label}
\end{figure*}

\noindent As shown in Figure~\ref{fig:enter-label}, the best performance was achieved with a window size of 64. This setting provides a trade-off between capturing fine-grained waveform details and maintaining sufficient temporal context for effective local-global feature integration.

\section{Conclusion and Future Work}

This study introduced LGA-ECG, a novel deep learning model for ECG classification that integrates local convolutional inductive biases with global self-attention mechanisms. Our approach effectively captures both fine-grained morphological features and broader temporal dependencies, leading to significant improvements over state-of-the-art methods. LGA-ECG achieved the highest F1-score among all evaluated models, demonstrating the benefits of local-global attention in medical signal analysis.

A promising and important future direction is extending LGA-ECG with self-supervised learning techniques to pretrain the model on large unlabeled ECG datasets before fine-tuning it for classification. This approach could enhance generalization and robustness, particularly for rare abnormalities with limited labeled data. Additionally, exploring domain adaptation methods may further improve model performance across diverse populations and recording settings, increasing its clinical applicability.

\section{Ethical Considerations}

While LGA-ECG demonstrates superior classification performance, its deployment in clinical settings must be approached with caution. AI-driven models should support, rather than replace, expert decision-making, ensuring that automated predictions are interpreted in the context of a professional medical evaluation. Additionally, to mitigate ethical concerns related to data privacy and patient confidentiality, all datasets used in this study are publicly available, properly anonymized, and handled following ethical guidelines. No personally identifiable information was accessible or used, ensuring compliance with data protection regulations while promoting responsible AI research in healthcare.

\section*{Acknowledgments}
This work is partially supported by CNPq, CAPES, Fapemig, as well as projects CIIA-Saúde and IAIA - INCT on AI.

\bibliographystyle{splncs04}
\bibliography{refs.bib}

\end{document}